\documentclass[openacc]{rstransa}

\usepackage{xcolor}
\usepackage{mathrsfs}
\usepackage[round]{natbib}

\titlehead{Research}

\begin{document}

\title{A follow-up strategy enabling discovery of electromagnetic counterparts to highly-magnified gravitationally-lensed gravitational waves}

\author{
Dan Ryczanowski$^{1,2}$,
Jeff Cooke$^{3,4}$, 
James Freeburn$^{3,4}$, 
Benjamin Gompertz$^{2}$, 
Christopher P. Haines$^{5}$, 
Matt Nicholl$^{6}$, 
Graham P. Smith$^{2,7}$, 
Natasha Van Bemmel$^{3,4}$, 
Jielai Zhang$^{4}$
on behalf of the multimessenger lensed-GW search project team
}

\address{Affiliations found at the end of the manuscript.}

\subject{xxxxx, xxxxx, xxxx}


\corres{Dan Ryczanowski\\
\email{dan.ryczanowski@port.ac.uk}}

\begin{abstract}
Making an unambiguous detection of lensed gravitational waves is challenging with current generation detectors due to large uncertainties in sky localisations and other inferred parameter distributions. However, in the case of binary neutron star (BNS) mergers this challenge can be overcome by detecting multiple images of its lensed kilonova counterpart, simultaneously confirming the lensing nature of the event and locating it precisely - further enabling a wealth of lensed multimessenger science. Such a strategy demands answers to two key problems: 1) How can candidate lensed BNS events be identified fast enough to ensure the lensed kilonova is still detectable? 2) What is the most economical observing strategy on telescope time for following up candidate lensed events to discover lensed kilonovae? This article will discuss solutions to both points, specifically: how GW detections of progenitors in the $\sim$ 2.5 to 5 $M_\odot$ black hole "mass gap" can be interpreted as candidate lensed BNS events, giving evidence for lensing from just a single detection, and will present a strategy that can actively be employed for follow-up of such events in the O4 run of LVK and beyond.
\end{abstract}


\begin{fmtext}

\end{fmtext}

\maketitle

\section{Introduction}

Confirmed detections of gravitationally-lensed gravitational waves (GWs) will mark historical stepping stones that enable advancements in astrophysics, particularly for detections of dark matter haloes and primordial black holes \citep[e.g.][]{Liao18,Oguri20,Diego20,Guo22,Urrutia22,Jana24} and progenitor population studies \citep[e.g.][]{Gais22,Xu22},
cosmology \citep[e.g.][]{Sereno11,Piorkowska13,Wei17,Ding19,Hannuksela20} and fundamental physics \citep[e.g.][]{Baker17,Collett17,Fan17} -- also see other publications in this issue. However, confirming the lensing nature of individual GW events is not as principally straightforward as for classical electromagnetic observations of lensed sources, where multiple images of a source can be individually resolved on the sky. 

In a GW detector, lensed events manifest as multiple strain signals emanating from the same source that each experience some level of lens magnification, and are separated by some time delay, where the scale of both factors is strongly dependent on the geometry of the source-lens system \citep[for a review on the fundamentals of lensed GWs, see][]{Grespan23}. Since each of these signals originates from the same source, the inferred parameters such as the masses, spins and redshifts/luminosity distances of the merging compact objects should be identical once corrected for magnification, and it is also expected the phase evolution for each GW in the detector will exactly match \citep{Dai20}. Furthermore, skymaps generated for the locations of the mergers should overlap, with the overlapping region containing the true location of the source. However, large uncertainties present in the posterior distributions of these quantities make comparisons difficult, and confirming a pair of events as lensed images of the same source with a high degree of confidence will likely require significantly more precise parameter estimation than is currently available.

Many searches for evidence of lensing in GWs in the literature \citep[e.g.][]{Abbott21,Goyal21,Li23,Lo23,Janquart23,Abbott24} focus principally on pairing lensed events from binary black hole (BBH) mergers using the GW strain signals. Agnostically, this is a sensible decision, since GWs detected from BBHs currently outnumber any other class of merger by essentially two orders of magnitude, and the relative rate of gravitationally lensed GW detections is expected to be $\sim$one per thousand detections that are not lensed \citep[][Smith et al. in this volume]{Xu22,Magare23,Smith23,Abbott24}. However, binary neutron star (BNS) mergers provide an additional channel of information via a kilonova (KN) counterpart that allows observers to simultaneously harness the strengths of both EM and GW signals. BBH mergers are not typically expected to produce direct EM counterparts, although AGN flares as a result of merger kicks are suggested candidates that may have already been discovered, but their correlation is unconfirmed \citep{Graham20}. Neutron star + black hole (NSBH) mergers may also produce KN-like EM counterparts, although these are predicted to be at least 2-3 mags fainter than those produced by BNS \citep[e.g.][]{Rosswog05,Tanaka14,Yang15,Kasen15,Gompertz23}. Therefore, detection of multiply-imaged KNe from a BNS merger would provide extremely robust evidence for lensing if one were discovered following a GW event, while also being the most feasible and most understood EM counterpart at present \citep[see e.g.][]{Nicholl24}.

Detection of lensed GWs is feasible today, in LVK's current fourth run (O4), although a large lens magnification $\mu$ is required to amplify the signal from the predicted high redshift ($z\simeq1-2$) population of BBHs and BNS to a level where it can be detected with current-generation detectors. \citet{Smith23} predict that for current detectors, BBH require $\mu \sim 10$ and BNS $\mu \sim 400$. In the next observing run O5, these required magnifications will decrease as detector sensitivity increases, to $\mu \sim 2-10$ for BBH and $\mu \sim 100$ for BNS. Drawing context from EM lensing observations, $\mu \simeq2-10$ is representative of common occurrences of multiple imaging by galaxy-scale lenses, meaning detection of a population of lensed GWs from BBHs becoming close to a reality. GWs from BNS will be still restricted to the high magnification regime, although there are known cases of extreme magnification of individual stars in the EM -- a striking example is the lensed star Earendel \citep{Welch22}, which holds the record for the most distant observed star at $z=6.2$, only detectable due to an extreme magnification of $\mu > 4000$. Even though a detection is unlikely, preparation is required to best capitalise on what would be a revolutionary once-in-a-century discovery. The aim of this paper is thus to present and motivate a search strategy for the follow-up of candidate highly-magnified lensed GW events from BNS mergers, in the search for their EM counterparts, relevant for current (O4) and future (O5) GW search runs. We begin with a short overview of lensed BNS rates in \autoref{sec:rates}, and describe how we use the  putative ``mass gap'' between neutron stars and black holes as a proxy for lensing of BNS mergers in GWs in \autoref{sec:mass_gap}. We finally review searches for EM counterparts to candidate lensed GWs in the literature, and describe the aforementioned strategy going forward in \autoref{sec:observing_strategies}.

\section{Lensed BNS rates}
\label{sec:rates}
Whilst detecting a lensed kilonova itself is no trivial feat, detections are feasible with current generation EM and GW observatories. The predicted rates for detecting lensed GWs from BNS mergers are approaching interesting and relevant values of $\lesssim 0.1$/year now in the current LVK run, O4, and $\lesssim 1$/year in the next observing run, O5 \citep{Smith23,Magare23,Abbott24}. These estimates are highly uncertain, with the dominant source of uncertainty coming from the local comoving rate density of BNS mergers $\mathscr{R}_0$, which is highly uncertain at $13 < \mathscr{R}_0 < 1900$ Gpc$^{-3}$yr$^{-1}$ \citep{Abbott23}. Recently, this rate has been updated by LVK using (the lack of) detections in O4a to $5 < \mathscr{R}_0 < 920$ Gpc$^{-3}$yr$^{-1}$. This information is not officially published by LVK, but details of this updated calculation can be found on their website \footnote{\url{https://emfollow.docs.ligo.org/userguide/capabilities.html}}. 

\section{Selecting candidate lensed GWs from BNS using the mass gap}
\label{sec:mass_gap}
Kilonovae are a class of rapidly-fading transients, implying that any follow up must be done quickly -- within a few hours to catch the rise, or within a few days in order to catch the object before it fades completely \citep{Villar2017,Abbott17,Metzger2019}. Therefore, in order to commence a search for a lensed KN we require a method that can promptly distinguish a lensed GW candidate and ideally not one that requires further time-consuming analysis or waiting for the arrival of a second image. The response time is especially short for a lensed kilonova, emerging from a higher redshift population (around $z\simeq1-2$) where the faster evolving rest-frame UV emission is redshifted into the observer-frame optical bands, compared to the slower evolving red emission observed from local KN populations (although there is some reprisal from time dilation effects). A proposed method that solves this issue is to consider GW events flagged in LVK alerts as comprising one or more compact objects within the ``mass gap'' as candidate lensed GWs. This is useful, since this mass information can be determined and distributed within minutes post-detection, and thus candidates can be selected from just a single LVK-detected `image` of the multiply-imaged lensed GW signal. The remainder of this section describes this method in more detail.

Because gravitational lensing amplifies the strain signal in the detector, when a lensed GW is detected the inferred masses are overestimated since the LVK GW models assume no lensing has taken place (i.e., $\mu = 1$). Therefore, a lensed BNS event can have its masses inferred to be erroneously within the mass gap -- a region of parameter space where observational evidence suggests there is a dearth of black holes -- existing approximately between 2.5 and 5 solar masses \citep{Farr11,Fryer12,Kreidberg12}. It should be noted that the boundary values of the mass gap are neither exact nor well defined, and its existence is currently uncertain -- especially since there have been a small number of black hole detections from LVK with mass estimates consistent with being in the mass gap \citep{LVK24}. However, the measurements are not precise enough to rule out the existence of the mass gap for certain. Furthermore, it is important to note that the existence of the mass gap is not a requirement for mass gap events to be interpreted as candidate lensed GWs. Its existence simply reduces significantly the number of non-lensed interlopers \citep{Smith23}. Theoretical models of compact object formation following stellar collapse can also support the existence of a mass gap \citep{Alsing18}, although this does not rule out the formation of these objects by other means, such as a direct merger.

Since O3, LVK have released public alerts in low latency to notify the community of potentially interesting events for follow-up. These alerts contain initial estimates for the masses, given as probabilities that the event is a specific type of progenitor: BBH, BNS, NSBH, mass-gap or terrestrial (noise). Originally, these probabilities summed to unity, but this was changed slightly in O4, removing the mass-gap probability from the main progenitor types in favour of a separate ``HasMassGap'' parameter. This joins ``HasNS'' and ``HasRemnant'' probabilities, which are useful parametrisations for follow-up observers. These signify the probability that the detected waveform was produced by at least one NS progenitor, or the probability that remnant matter remains after the merger in the form of dynamical or tidal ejecta, respectively. See \citet{Chaudhary24} for a detailed summary of the O4 alert strategy. These are released along with the primary progenitor probabilities and a skymap of the localisation typically minutes after a GW trigger. As we will later explore, the HasMassGap probability and the size of the localisation are the key parameters for lensed GW follow-up. 

Models predict 70 per cent of lensed BNS will have measured masses within the mass gap class \citep{Smith23}, leading to a classification with a high HasMassGap probability (hereafter we refer to these as `mass-gap events'). The remainder will spill mostly into BBH classifications and will require further information to be coaxed out from the non-lensed population. It should be noted that LVK's classifications can change between the initial prompt human-vetted analyses and the final catalogue-published masses. In \citet{Bianconi23}, EM follow-up was conducted following mass-gap events, triggering based on the low-latency human-vetted analyses, but these were eventually re-classified as non-mass-gap in \citet{Abbott23}. This highlights the importance of having accurate and reliable classifications from LVK in low latency to ensure telescope time is used efficiently. Furthermore, the release of additional GW information in low latency to the community would provide additional context to help identify false positives. For example, knowing the mass ratio of the merger would help rule out cases of non-lensed high mass ratio BBH and NSBH, since (lensed) BNS would be expected to have mass ratios closer to unity. Importantly, mass ratio is invariant to lens magnification.

\section{Observing Strategies for Lensed GW Followup}
\label{sec:observing_strategies}
\subsection{Overview of various strategies}
\label{sec:strat_overview}
The follow-up of candidate lensed GW events has subtle differences from typical GW follow-up in terms of the employed strategy, but both share an end goal of detecting the EM transient associated with a recent GW trigger from LVK. For a comprehensive review of GW follow-up in other EM contexts, see the recent review from \citet{Nicholl24}. A very brief summary is that typical BNS follow-up either observes the highest probability regions of the GW sky localisation with large FoV telescopes, or targets specific galaxies using narrower-field telescopes -- those consistent with GW distance estimates or those with large stellar mass content within and near the localisation's peak. For lensed GW follow-up, we operate under the hypothesis that the signal is lensed, meaning in theory one only needs to check the locations of lensed lines of sight. However these lines of sight are so numerous within a GW sky localisation \citep{Robertson20}, and the cataloguing of them is so incomplete \citep{Ryczanowski20}, that a wide-field search is the most efficient use of telescope time for these searches. This motivates the evolution of programmes from pilot observations of massive galaxy cluster lenses in O2 and O3 (\citet{Smith19,Bianconi23}; analogous to pointed observations of individual massive galaxies in ``traditional'' GW follow-up programmes) to wide-field lensing specific follow-up observations in O4 and future runs, as described here.

The most time efficient solution to ensure full coverage of the entire lensing cross-section within a GW localisation is therefore to image the entire localisation to the required sensitivity, and as soon as possible after detection of a candidate lensed event, and to repeat this on subsequent nights. (We discuss this type of strategy in more detail in \autoref{sec:present_strat} as a feasible and current method.) The size of the sky localisation is the clear limiting factor on the depth achieved in the above strategy, and localisation precision is heavily improved by a greater number of interferometers detecting the same event. Detection with at least three detectors makes a great difference compared to two (or fewer) detectors, meaning the greater the number of detectors in the network vastly increases the strategy's feasibility. Ideally, the network consists of even more $(>3)$ detectors, since individual detectors often have periods of required maintenance and other downtime during which they do not collect data. In context of the most recent O4 observing run, Virgo was offline or operating at limited range for most of the first half, so localisations were too large to enable this kind of search.

Strategies in the future can lean on wide-field surveys such as Rubin's LSST in order to search for candidate counterparts to (lensed) GW events. These have the advantage that they are regularly scanning large fields irrespective of GW detection, and discover (lensed) KNe independently -- even when LVK's detectors are offline. The wide-fast-deep component of Rubin's LSST is sensitive to approximately the same magnifications of lensed KNe as the population of lensed GWs LVK will detect at forecasted O5 sensitivity (Smith et al., this volume). This means whilst Rubin has the capability to make lensed KN discoveries during survey mode, GW sky localisations will not necessarily be available for all such events, especially considering the gap between O4 and O5. As such, triggering dedicated follow-up in the localisation of candidate lensed BNS is strongly favoured to ensure deep observations are taken promptly after such a detection in order to catch the EM counterpart. That being said, discoveries of lensed KNe within the alert stream of Rubin are certainly possible, and would benefit from searches specifically targeting rapidly-fading transients, such as ZTFReST \citep{Andreoni21}, which was applied to ZTF data to search for (unlensed) kilonovae. 

Targeted lensed GW follow-up will also be different, and have greater prospects in the Rubin era. Rubin itself will be the most powerful machine for follow-up of GW localisations owing to its wide FoV and depth. A few per cent of Rubin's time will be dedicated for exceptional target-of-opportunity (ToO) observations. The rarity of lensed GWs means any such follow-up efforts are very undemanding on telescope time when integrated over the entire survey, and are thus a prime candidate. Recommendations in \citet{Andreoni24} report 0.4\% of Rubin's observing time is sufficient to follow up all lensed GW candidates in O5, using a strategy akin to the one following in \autoref{sec:present_strat}.

\subsection{Present-era discovery and follow-up strategy}
\label{sec:present_strat}
\begin{figure}
    \centering
    \includegraphics[width=1.0\textwidth]{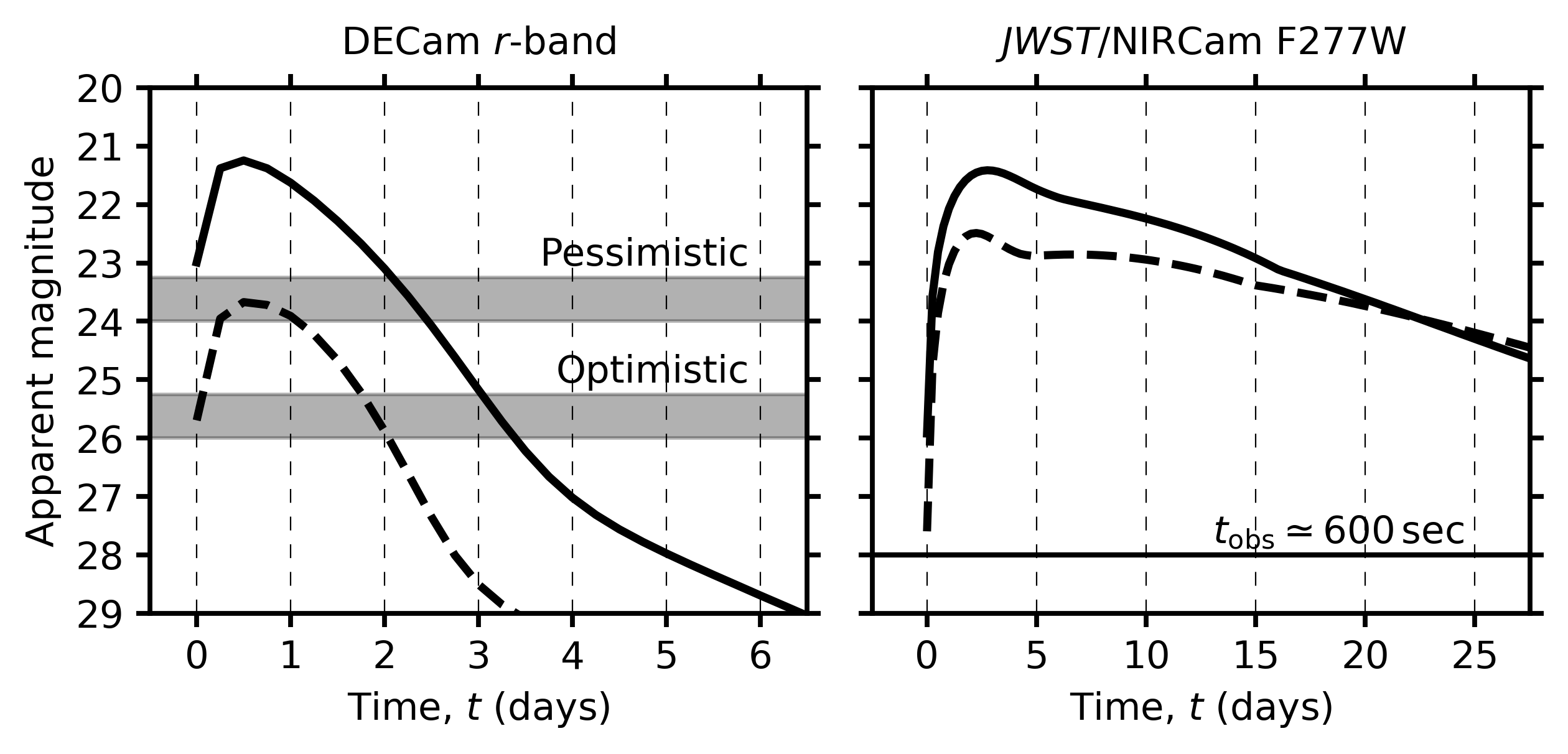}
    \caption{{\bf Left:} The predicted observer-frame $r$-band lightcurves of lensed KN counterparts to lensed BNS detectable by current generation GW detectors. The solid lightcurve shows a model that reproduces AT2017gfo, the KN counterpart to GW170817 -- the only GW with a detected counterpart to date. The dashed lightcurve shows a redder, more conservative model in which the observer has a less favourable viewing angle than for AT2017gfo and the early blue KN emission is suppressed. The 5$\sigma$ point source sensitivities achievable with DECam in `pessimistic'/poor conditions and in `optimistic'/ideal conditions (see text) are shown as horizontal bands. The horizontal grey bands represent how the achievable depths vary with size of sky localisation, given a fixed half night of observing time: the bottom edge representing 50 deg$^2$ and the top edge 200 deg$^2$. Even for 200 deg$^2$ regions such observations are still sensitive to AT2017gfo-like counterparts in suboptimal conditions. {\bf Right:} Predicted {\it JWST}/NIRCam F277W lightcurve showing detectability of lensed KNe using the same models as on the left. The solid horizontal line highlights that targeted follow-up observations in the IR are feasible for $\sim$1 month post-event. {\it JWST} is shown since it is the only feasible instrument for spectroscopic confirmation and accurate astrometry.
}
    \label{fig:fig}
\end{figure}

With detections from current-era GW and current-era EM observatories (defined as the facilities available during O4), the best strategy from a practical standpoint uses dedicated follow up to cover as much of the localisation of mass-gap events as possible, and to go as deep as possible in the $\sim$night following detection. This ToO strategy ensures coverage of the full strong-lensing cross-section within GW localisations whilst enabling observations to begin fast enough to capture lensed KNe at their peak. The remainder of this section will go into further detail on the strategy, provide estimates for how deep such observations can (and are required to) go as a function of sky localisation size, and other specific choices for which events are suitable for follow-up.

Based on recent models, lensed BNS systems at $z \simeq 1-2$ are detectable with current-generation GW interferometers provided they are magnified by $\mu\sim100$ \citep{Smith23,Magare23}. Using the KN models of \citet{Nicholl21}, we simulate lightcurves for this population of sources to estimate their detectability under various observational setups and conditions, and to determine how much observing time needs to be invested to make discoveries of lensed KN counterparts feasible. \autoref{fig:fig} shows example model lightcurves for these lensed KNe with 5$\sigma$ observing limits for DECam on the Blanco telescope, marked with the horizontal grey bands. DECam is chosen as it is the most powerful instrument (in terms of depth and FoV) for lensed KN follow-up pre-Rubin, when utilising the selection strategy described in \autoref{sec:false_positives}. Based on these lightcurves, we tune an integration time for these observations so that even for a conservative KN model with a less favourable viewing angle (dashed curve), the KN is detectable at peak in a pessimistic scenario with poor observing conditions (full moon, low elevation, poor seeing and some cloud extinction). This is marked by the bottom edge of the ``pessimistic'' labelled grey band, at $\sim24$ mag. In addition, based on these models, for best prospects of catching the lensed KN at/around peak observations should no later than 36 hours post GW trigger. This essentially limits any follow-up to the night immediately following GW detection, and we therefore assume half a night as the ceiling for the total available useful follow-up time, such that the observability of target fields is always accounted for regardless of declination. Based on the above conditions, approximately 20 pointings reaching 24 mag can reasonably be made in half a night, which leads to a nightly coverage of about 50 square degrees of sky for a telescope with DECam's 3 sq degree field-of-view. This effectively limits GW detections to be localised more precisely than 50 square degrees in order to be suitable for follow-up. It should be noted, however, that 50 sq degrees is representative of the smallest $\sim$ 10 per cent of predictions for GW skymaps \citep{Petrov22,Kiendrebeogo23}. Therefore, it is useful to consider flexibility in follow-up campaigns, i.e., adjusting the individual exposure time to allow for additional exposures per night and enabling wider coverage. The top edge of the grey bands represents this flexible scenario where we have used shorter individual exposures and sacrificed some sensitivity, but can hence extend the number of pointings fourfold up to approximately 200 sq. degrees whilst only no longer being sensitive to redder kilonova models with unfavourable viewing angles in pessimistic observing scenarios. Sensitivity to KN physics in optimistic conditions remains essentially unchanged. Further adjustments can be made trading depth for coverage, but three-detector networks are well capable of producing GW skymaps below 200 sq. degrees and this would risk becoming insensitive to a significantly larger proportion of the lensed KN population.

As mentioned in \autoref{sec:strat_overview}, Rubin will provide a substantial boost to the feasibility of ToO follow-up observations for lensed GWs. We refer the reader to \citet{Andreoni24} for a comprehensive report, but summarise the main conclusions here. Retaining the requirement that each epoch of ToO follow-up be completed in within half a night, Rubin is capable of following up an AT2017gfo-like counterpart to a lensed BNS event for sky localisations as large as 900 sq. degrees, and redder KNe with less favourable viewing angles for only the best-localised events of 15 sq. degrees. These estimates are determined using the same underlying models as the solid and dashed curves in \autoref{fig:fig} respectively, so it is tempting to directly compare these numbers to the 200 and 50 sq. degree regions determined previously for DECam -- however, these are not relatable quantities. Follow-up efforts are driven by detections of lensed GWs, for which the observed population is determined by detector sensitivity in LVK's O5 which overlaps with Rubin -- the median detected source luminosity distance is forecasted to increase to 740 Mpc from 400 Mpc in O4 \footnote{\url{https://emfollow.docs.ligo.org/userguide/capabilities.html}} (used in the DECam estimations). The net result of this is that the magnification required for lensed BNS GW detection from the same underlying source population drops from $\mu\sim400$ to $\mu\sim100$ \citep{Smith23}. As a result, the bulk of the detectable lensed BNS GW population is now lower magnification, and hence their lensed KNe counterparts are also fainter. The relative rate of possible detections increases by $\sim1-2$ orders of magnitude \citep{Smith23}, but at the cost of needing to probe greater depths to make these detections. Rubin's enhanced capability handles these extra constraints well for the brighter models, and even provides flexibility for obtaining multiple bands of photometry (discussed further in \autoref{sec:false_positives}). Similar to DECam in O4 however, Rubin is still restricted to only the best localised events to detect fainter KNe.

These observations are also unique in that they are among, if not the, deepest follow-up observations of GW sources. DECam observations can reach 26th $r$-band magnitude in optimal conditions, whereas Rubin can extend this to 27th in $r$ and 27.5 in $g$. This is to accommodate all models of lensed KNe as stated above, including those that are redder, fainter or viewed off-axis compared to an AT2017gfo-like model. It is also one of few follow-up programmes that trigger on mass-gap events, and could facilitate interesting and unexpected discoveries of optical counterparts in these regimes.

\subsection{Identifying lensed KNe and false positives}
\label{sec:false_positives}
KNe are thought to be unique compared to other common classes of explosive transient. Their intrinsic luminosities $M \sim -11$ to $14$ mag sit between novae ($M > -10$ mag) and supernovae ($M < -14$), and they fade much more rapidly than typical supernovae in observer-frame optical bands. As previously mentioned, lensed KNe fade even quicker, because the faster-evolving rest-frame UV emission is redshifted into observer-frame optical bands. This is in contrast to KNe that are not lensed, which are detectable at lower redshift and the rest-frame optical probes an entirely different regime of comparatively slower-evolving red emission. This rapid fading is the key to selecting a lensed KN among an abundance of other transient detections, of which many are expected in such deep imaging. Among the other classes of transient, few fade as fast as lensed KNe. ``Fast blue optical transients'' (FBOTs), are close in terms of decline rate, decaying by 2 mags on timescales of $\sim 1$ week \citep{Prentice18} -- but are still noticeably slower than the lightcurve in \autoref{fig:fig}. Cataclysmic Variables (CVs) or other fast-fading novae are known to fade as rapidly as lensed KNe, but these are in the fast tail-end of that distribution. Since these are intrinsically very faint objects they will likely only be contaminants close to the galactic plane or within nearby galaxies. False positives can further be suppressed by the expectation that a true lensed KN will be discovered in the proximity of a lens, and in some cases a lensed host galaxy. Typical lenses consist of red early-type galaxies and galaxy clusters that should easily be visible in deep imaging. Therefore, requiring coincidence with a plausible lens can help rule out false positives, particularly those within our own galaxy. A lensed KN host galaxy may not always be visible due to its intrinsic properties, and the host galaxies of lensed transients are not always themselves multiply-imaged or as highly magnified \citep[see e.g.][]{Sainzdemurieta24}, especially when considering many KNe are expected to have large separations from their host galaxy \citep[see e.g.][and references therein]{Gaspari24}. Therefore observing the lensed host is a possible diagnostic, but not in all cases.

Due to the lack of false positives in the brightness and decline rate parameter space as described, we opt to utilise a selection for lensed KNe based solely on how fast a transient fades. We note that this can be achieved by taking three observing epochs of the entire field. The second of the three should be taken ideally the night after the first, or otherwise as soon as possible, and is used specifically for the selection of candidates: subtracting the second epoch from the first can extract any point sources that have faded by a significant amount ($>1$ mag). The third epoch is then used as a post-hoc template image and can be taken flexibly some time after any KNe should have completely faded ($t \gtrsim 4$ days post GW trigger). This third image will also allow for accurate host-subtraction and hence provide accurate flux measurements of a putative KN detection, and can help to rule out some false positives such as variable stars that may show a re-brightening.

An advantage of the selection technique described above is that it can be performed using just a single photometric band. This synergises well with the observing time restrictions of the first epoch, since observations can focus on the key aspects of maximising depth and coverage in a single band rather than distributing time across multiple. This strategy is also an application for wide-band filters, such as DECam's $VR$ filter and the $L$ filter on the GOTO telescope array \citep{Steeghs22}, where the priority is to maximise number of collected photons rather than to discern colour or wavelength information. Future follow-up with Rubin may be flexible enough to permit more than one filter to observe the entire localisation within the one night window, owing to its wide FoV. This colour information can help to further constrain candidates to those consistent with KNe \citep{Andreoni24}.

Other than the false positives mentioned above that can be ruled out by rate of fading, some other common false positives can persist through our selection. Among these are rapidly-oscillating variable stars and quasars. A third observational epoch may help constrain these, or otherwise they can be removed by cross-matching with deep catalogues \citep[e.g.][]{Drlica-Wagner22}. Catalogues do not, however, exist equally in depth in all areas of the sky, so some regions may be more prone to contamination by quasars and variable stars than others. This is especially true since these observations will be very deep, and hence may explore uncharted depths below $m\sim25$mag. These observations also have the opportunity to discover rarer and scientifically valuable transients. This includes KNe that are not lensed, which may or may not be related to the GW event, for which detailed observations are likely to be scientifically valuable nonetheless. To put this in context, the ZTFReST search for fast-fading transients in the ZTF archival data and alert stream \citep{Andreoni21} reported 13 transients consistent with kilonova fading rates ($>0.3$ mag/day) in their 13 months of science validation. 5 of these 13 are consistent with lensed kilonova fading rates ($>1$ mag/day), and all are reported as GRB afterglows. This demonstrates 1) the low intrinsic rates of $>1$ mag/day fading transients which may act as false positives to lensed KNe, and 2) that those objects are scientifically interesting despite not being lensed KNe. In a similar vein, discovery of new fast transients is also possible, similar to the serendipitous detection of HFF14Spo \citep{Rodney18}: a highly-magnified one-of-a-kind event discovered in a gravitationally-lensed host galaxy.

The final step is to use targeted space-based follow-up of candidates discovered in the wide-field imaging, which is essential for two reasons. Firstly, unless the seeing is good and the image separation of the lensed KN is large, the multiple images will not be resolvable from the ground. In any case, accurate space-based astrometry is required to control uncertainties in $H_0$ constraints from multi-messenger time delay cosmography (Birrer et al., this volume). Secondly, the rapid fading of a high-redshift KN in optical bands means it is essentially unobservable after only a few days post-peak in rest-frame optical bands, even in the most optimal scenario (\autoref{fig:fig}, left panel). However, infra-red (IR) observations, either from space or the ground, are expected to significantly prolong the observing window. IR instruments are not optimal for initial detection because their FoV is smaller then (say) DECam, so candidates must be determined in the optical beforehand. Space-based follow-up is preferred, however, due to the increased resolution without contention from the atmosphere as well as for increased depths for fainter and less-magnified cases in the Rubin era. The right-hand plot of \autoref{fig:fig} shows the same model KNe lightcurves from the left panel, except in {\it JWST}/NIRCam's F277W filter. Due to the inherent sensitivity of {\it JWST} and the rest frame wavelengths it probes at high-z, lensed KNe are detectable for a much longer period than in optical bands from the ground. As shown in the figure, a 600 second exposure can detect a lensed KN for over a month post-explosion, meaning the rapid turn around times required for initial discovery are not constraining for detailed follow up of the system. 

\section{Conclusions}
With current generation GW detectors, detecting a multiply-imaged KN in the wake of a GW signal may be the only way to robustly confirm that the GW was gravitationally lensed. As such, this article outlined how GWs with masses inferred in low latency to be within the range of the mass gap of $\sim$2.5--5 M$_\odot$ are candidate lensed GWs from BNS systems. A strategy to follow up such candidate lensed events with wide field-of-view optical such as DECam on Blanco and LSSTCam on Rubin was then outlined, and motivated using recent models of lensed KNe lightcurves. The strategy involves covering as much of a GW sky localisation as possible in half a night and going as deep as possible in that night immediately following the GW trigger, before repeating observations within a few days of the first in order to pick out new point sources that are fading very rapidly. A third epoch of images is then taken after any prospective KNe have faded to act as a post-hoc template. Promising candidates that are consistent with a lensed KN are subsequently targeted by space-based instruments for detailed follow-up, in order to resolve the multiple images, monitor post-peak evolution and provide crucial spectroscopic confirmation. Such a strategy is designed to be optimal for demanding as little telescope time as possible, and yet harbours great potential to make a significant discovery for multimessenger astronomy.

\ack{The authors would like to extend their thanks to the organisers of the Royal Society Specialist Discussion Meeting, ‘Multimessenger Gravitational Lensing’, as well as to Richard Massey, Armin Rest and Frank Valdes for stimulating discussions. We also thank the two anonymous reviewers, whose comments helped greatly enhance this manuscript. DR is supported by the European Research Council (ERC) under the European Union’s Horizon 2020 research and innovation programme (LensEra: grant agreement No 945536). GPS acknowledges support from The Royal Society, the Leverhulme Trust, and the Science and Technology Facilities Council (grant number ST/X001296/1).}\\
\\
{\bf Affiliations:}\\
\noindent $^{1}$Institute of Cosmology and Gravitation, University of Portsmouth, Burnaby Rd, Portsmouth, PO1 3FX, UK\\
$^{2}$School of Physics and Astronomy, University of Birmingham, Edgbaston, Birmingham B15 2TT, UK\\
$^{3}$Centre for Astrophysics and Supercomputing, Swinburne University of Technology, Hawthorn, VIC 3122, Australia\\
$^{4}$ARC Centre of Excellence for Gravitational Wave Discovery (OzGrav), Hawthorn, VIC 3122, Australia\\
$^{5}$Instituto de Astronomía y Ciencias Planetarias de Atacama (INCT),
Universidad de Atacama, Copayapu 485, Copiapó, Chile\\
$^{6}$Astrophysics Research Centre, School of Mathematics and Physics, Queens University Belfast, Belfast BT7 1NN, UK\\
$^7$Department of Astrophysics, University of Vienna, T\"urkenschanzstrasse 17, 1180, Vienna, Austria

\bibliographystyle{abbrvnat}
\bibliography{sample.bib}

\end{document}